\documentclass[twocolumn]{aastex63}

\hypersetup{linkcolor=red,citecolor=blue,filecolor=cyan,urlcolor=blue}
\usepackage{graphicx,color}
\usepackage{amssymb}
\usepackage{amsmath}
\usepackage{url}
\usepackage{natbib}
\usepackage{txfonts}
\usepackage{multirow}
\usepackage{array}
\usepackage{rotating}
\usepackage{sidecap}
\usepackage{hyperref}
\usepackage{epstopdf}
\usepackage{footnote}
\usepackage{tabularx}
\usepackage{booktabs}
\usepackage{longtable}
\usepackage{tabu}
\usepackage{longtable}

\newcommand{\SiIV}{\ion{Si}{4}}
\newcommand{\CII}{\ion{C}{2}}
\newcommand{\MgII}{\ion{Mg}{2}}

\newcommand{\Halpha}{H$\alpha$}

\newcommand{\kms}{km~s$^{-1}$}

\newcommand{\sdo}{\textit{SDO}}
\newcommand{\iris}{\textit{IRIS}}

\shortauthors{Panesar et al.}

\shorttitle{Counter-streaming Flows Driven by Network Jets}
\shortauthors{Panesar et al.}

\begin{document}
	\title{Network Jets as the Driver of Counter-Streaming Flows in a Solar Filament/Filament Channel}

\correspondingauthor{Navdeep K. Panesar}
\email{panesar@lmsal.com}

\author[0000-0001-7620-362X]{Navdeep K. Panesar}

\affil{Lockheed Martin Solar and Astrophysics Laboratory, 3251 Hanover Street, Bldg. 252, Palo Alto, CA 94304, USA}
\affil{Bay Area Environmental Research Institute, NASA Research Park, Moffett Field, CA 94035, USA}

	\author[0000-0001-7817-2978]{Sanjiv K. Tiwari}
\affil{Lockheed Martin Solar and Astrophysics Laboratory, 3251 Hanover Street, Bldg. 252, Palo Alto, CA 94304, USA}
\affil{Bay Area Environmental Research Institute, NASA Research Park, Moffett Field, CA 94035, USA}

\author[0000-0002-5691-6152]{Ronald L. Moore}
\affiliation{Center for Space Plasma and Aeronomic Research (CSPAR), UAH, Huntsville, AL 35805, USA}
\affiliation{NASA Marshall Space Flight Center, Huntsville, AL 35812, USA}

\author[0000-0003-1281-897X]{Alphonse C. Sterling}
\affiliation{NASA Marshall Space Flight Center, Huntsville, AL 35812, USA}


\begin{abstract}

Counter-streaming flows in a small (100\arcsec-long) solar filament/filament channel are directly observed in high-resolution \sdo/AIA EUV images of a region of enhanced magnetic network. We combine images from \sdo/AIA, \sdo/HMI and \iris\ to investigate the driving mechanism of these flows. We find that: (i)  counter-streaming  flows are present along adjacent filament/filament channel threads for $\sim$ 2 hours, (ii) both ends of the filament/filament channel are rooted at the edges of  magnetic network flux lanes along which there are impinging fine-scale opposite-polarity flux patches, (iii) recurrent small-scale jets (known as \textit{network jets}) occur at the edges of the magnetic network flux lanes at the ends of the filament/filament channel, (iv) the recurrent network jet eruptions clearly drive the counter-streaming flows along threads of the filament/filament channel, (v) some of the  network jets appear to stem from sites of  flux cancelation, between  network flux and merging opposite-polarity flux, and (vi) some show brightening at their bases, analogous to the base brightening in coronal jets. The average speed of the counter-streaming flows along the filament/filament channel threads is 70 \kms. The average widths of the AIA filament/filament channel and the \Halpha\ filament are 4\arcsec\ and 2.5\arcsec, respectively, consistent with the earlier findings that filaments in EUV images are wider than  in \Halpha\ images. Thus, our observations show that the continually repeated counter-streaming flows come from network jets, and these driving network-jet eruptions are possibly prepared and triggered by magnetic flux cancelation. 

\end{abstract}

\keywords{Sun: Filament --- Sun: chromosphere---  Sun: corona --- Sun: magnetic fields }

\section{Introduction} \label{sec:intro}

Solar filaments (on-disk), and/or solar prominences (off-limb) are mainly composed of cool and dense plasma,  suspended in the  hotter solar corona  \citep{eng76,hir85,tand95,mackay10,lab10,parenti14}. They lie above photospheric magnetic neutral lines (also known as polarity inversion lines; \citealt{bab55}), i.e.  between opposite-polarity magnetic field regions \citep{mar73}. The cool filament plasma is trapped in highly sheared magnetic field lines \citep{balle90,martens01}.

It is well known that the macroscopic filament structure is composed of several fine-scale  threads \citep{eng76,engvold01,lin05b}. The main body of a filament, known as the spine, runs horizontally along the photospheric neutral line. So-called barbs protrude from both sides of the spine and connect to the photosphere \citep{mar98}; they  supposedly  tether the filament system \citep{eng76}. The presence of cool plasma flows along the spine of the filament, as well as in the barbs, have been observed in high resolution images \citep{lin03,lin05b}. Sometimes  plasma flows in opposite directions (bi-directional flows)  are seen along a filament thread or along adjacent filament threads.

Using H$\alpha$ data from BBSO, \cite{zir98} reported bi-directional/anti-parallel flows, which they named as  ``counter-streaming" flows, in the entire  structure of the filament (including spine and barbs). They detected both redshifts and blueshifts in the wings of the H$\alpha$ line, which indicate  plasma flows toward and away from the observer. In their study, what drives the bi-directional flows remains unknown. Later, similar bi-directional flows were also reported by \cite{lin03,lin05b,chae07,schmieder08,panasenco08} and \cite{ahn10}.

\cite{alexander13}  reported counter-streaming flows in an active region filament using high resolution EUV 193 \AA\ images from the High-resolution Coronal Imager (\textit{Hi-C}) and they concluded that  counter-streaming flows are not discernible  in the \textit{Solar Dynamics Observatory} (\sdo)/Atmospheric Imaging Assembly (AIA) data because of its larger pixel size in comparison to Hi-C. Recently, \cite{diercke18}  reported anti-parallel flows in a quiescent filament using EUV images from  \sdo/AIA. However, the flows were not directly visible in the AIA images; \cite{diercke18} had to apply image enhancement techniques and local correlation tracking (LCT) to the AIA images to detect the horizontal flows along the filament. The driving mechanism of the bi-directional flows are not  investigated in any of the above mentioned studies.

A possible driving mechanism for the flows is magnetic reconnection at the ends of the filament. But until now, there have been no systematic observational studies of the origin of the bi-directional flows in filaments. In this Letter, we  address this key question: \textit{What drives the counter-streaming flows in solar filaments/filament channels}.

Recently, \cite{panesar18b,panesar19} reported small-scale network jets (such as observed by \citealt{tian14} and also known as \textit{jetlets} \citealt{raouafi14}) that are proposed to be miniature versions of coronal jets, because they: (i) occur at the edges of magnetic network lanes between majority-polarity network flux and merging minority-polarity flux, (ii) show brightenings at their base, and (iii) are located at the feet of far-reaching coronal magnetic field. The magnetic flux cancelation between majority-polarity network flux and merging minority-polarity magnetic flux prepares and triggers the jetlet eruption that ejects plasma out along the pre-existing ambient far-reaching field, i.e., makes the jetlet spire. All above properties are observed in most coronal jets (see e.g., \citealt{panesar16b,mcglasson19}, and references therein).

Here, from high resolution observations from \sdo\ and \iris\ we report counter-streaming plasma flows in a filament/filament channel and that they are driven by  network jets. To the best of our knowledge, this is the first report of direct \sdo/AIA observations of counter-streaming flows in a filament/filament channel displayed without using any image enhancement technique. By studying  continually repeated counter-streaming flows for two hours in the filament channel, we find that counter-streaming flows come from repeated jetlets  at both ends of the filament/filament channel. These network jets occur at sites of flux cancelation at edges of magnetic network flux lanes. We use the terms `network jets' and `jetlets' interchangeably throughout the paper. On the basis of the similarity of the observed flows in our filament with those in previous works \citep[e.g.,][]{zir98,alexander13,diercke18}, and to keep consistency with the terminology in these papers, here we mostly use the term ``counter-streaming" to describe the anti-parallel/bi-directional flows.” 

\begin{figure}
	\centering
	\includegraphics[width=\linewidth]{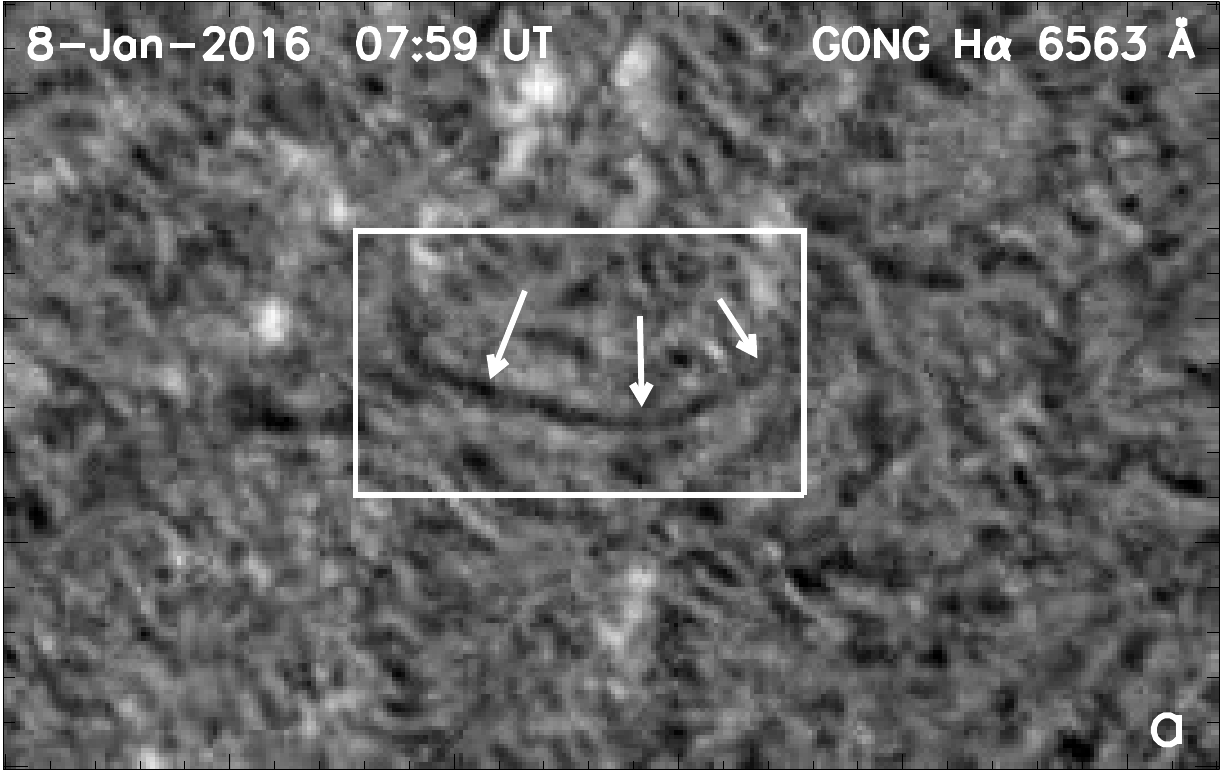}\vspace{0.3cm}
	\includegraphics[width=\linewidth]{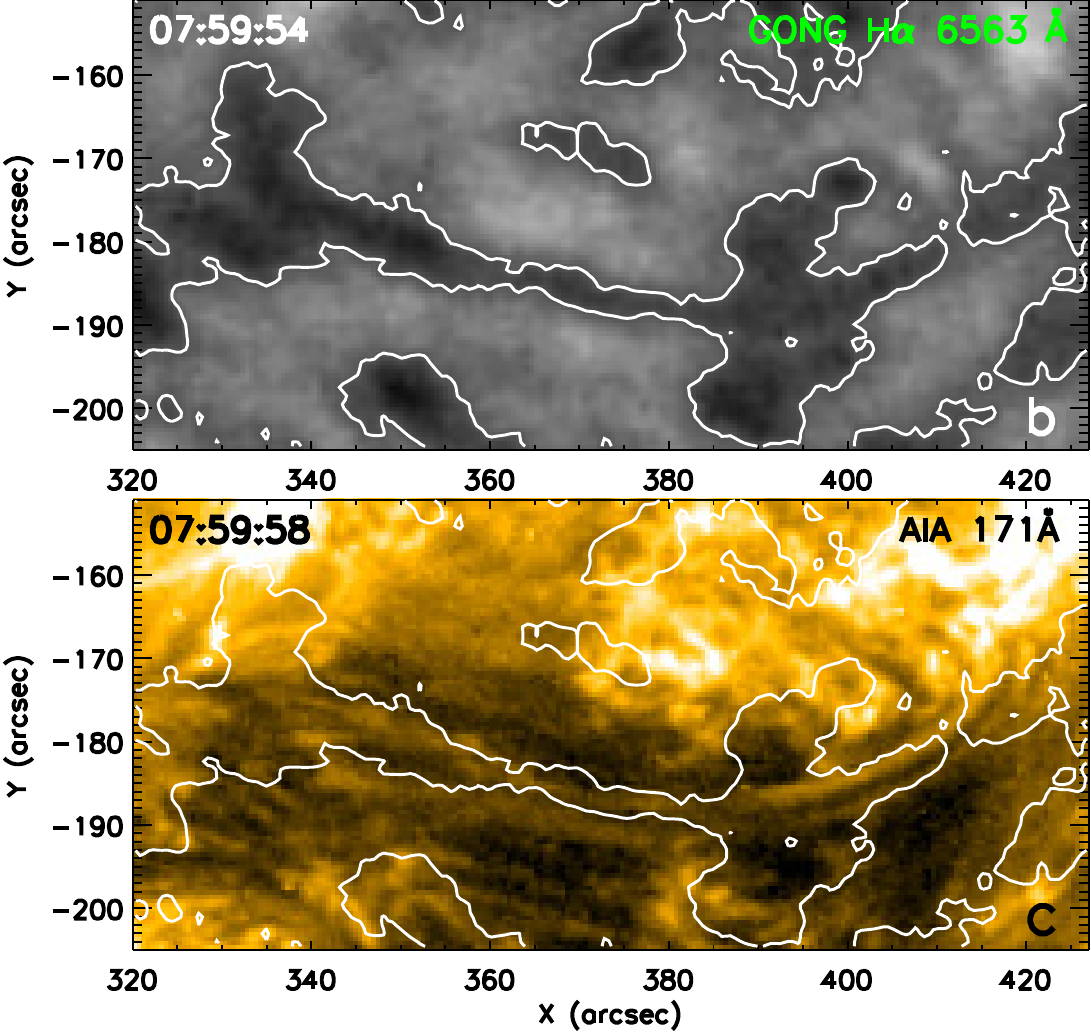} 
	\caption{A small solar filament/filament channel observed on 8-Jan-2016. Panel (a) shows the GONG H$\alpha$ image of the solar filament. The white arrows point to the filament and the white box shows the field of view analyzed in detail, and shown in Figures \ref{fig1h}(b, c), and Figure \ref{fig1}. Panels (b) and (c) show a darkness contour of the GONG \Halpha\ image of the dark filament drawn on a co-temporal AIA 171 \AA\ image of the same field of view.  That is, the white contour in (c) is from the GONG \Halpha\ image in (b) (at 07:59:54 UT) and outlines the filament/filament channel in the GONG \Halpha\ image.}  \label{fig1h}
\end{figure} 

\floattable
\begin{center}
	
	\begin{table}
		
		\setlength{\tabcolsep}{8pt} 
		
		\scriptsize{
			\caption{Key Data of  Jetlets at the Ends of the filament/filament channel \label{tab:list}}
			\renewcommand{\arraystretch}{1.2}
			\begin{tabular}{c*{5}{c}}
				\noalign{\smallskip}\tableline\tableline \noalign{\smallskip}
Event & East End\tablenotemark{a} &   Base\tablenotemark{b}   & Discernible\tablenotemark{c} & IRIS\tablenotemark{d} \\
No. & Jetlet Timing   &   Brightening &  Minority flux &  Coverage\\
				
\noalign{\smallskip}\hline \noalign{\smallskip}

1. & 07:36:34  & No\tablenotemark{e} & Yes & No\\ 
				
2. & 07:59:25  & Yes & Yes & Yes\\ 
				
				
3. & 08:21:22   & Yes & Yes & Yes\\ 
				
				
4. & 08:38:01\tablenotemark{f}   & Yes &  Yes?\tablenotemark{g}  & Yes\\ 
				
				
5. & 08:42:16   & Yes &  Yes & Yes\\
				

6. & 08:44:23   & Yes &  Yes & Yes\\

				
\noalign{\smallskip}\tableline\tableline \noalign{\smallskip}
& West End\tablenotemark{h} &   Base\tablenotemark{i}\\
 & Jetlet Timing  &   Brightening  \\
\hline
7. & 07:14:10  & Yes & Yes &  No \\ 
				
8. & 07:37:22  & Yes & Yes &  No \\ 
				
9. & 07:45:22  & No & Yes &  No \\ 
				
10. & 08:43:46  & No & Yes?\tablenotemark{g} &  No \\ 
				
\hline
				
\end{tabular}
			
		\tablenotetext{a}{Timing of the visibility of the jetlet spire at the east end of the filament using \iris\ \SiIV\ SJIs. }
		\tablenotetext{b}{Whether a brightening at the base of the jetlet is discernible in \iris\ \SiIV\  SJIs.} 
		\tablenotetext{c}{Whether minority-polarity flux is discernible at the base of the jetlets.}
		\tablenotetext{d}{Whether this jetlet is observed by \iris.}
		\tablenotetext{e}{This jetlet occurs before the start of the \iris\  observations (07:50:33--08:47:33).}
		\tablenotetext{f}{\iris\ has some bad frames (due to south atlantic anomaly) so the jetlet is not visible at this time. Therefore, we use AIA 171 \AA\ images for timing of this event.}
		\tablenotetext{g}{Minority-polarity flux is visible near the base of the jetlet but is not as near as in the other eight events.}
		\tablenotetext{h}{Timing of the visibility of the jetlet spire  at the west end of the filament using AIA 171 images (Figure \ref{fig3}).}
		\tablenotetext{i}{Whether a brightening at the base is discernible in AIA 171 \AA.} 
		}
		
	\end{table}
	
\end{center}

\section{DATA SET}\label{data}

We used data from the \textit{Interface Region Imaging Spectrograph} (\iris; \citealt{pontieu14}),  AIA \citep{lem12} and Helioseismic and Magnetic Imager (HMI; \citealt{scherrer12}) from \sdo\ \citep{pensell12}, and \textit{Global Oscillation Network Group} (\textit{GONG}; \citealt{harvey96}). A small filament (100\arcsec-long)  was observed in a region of enhanced magnetic network on the solar disk (330\arcsec, -175\arcsec) on 8-Jan-2016 by \iris\ and \sdo/AIA and HMI. \iris\  captured the filament/filament channel for about an hour, from 07:50 -- 08:47 UT. We study the same region for 2 hours (07:00--09:00) using \sdo\ data.

IRIS provides both slit-jaw images (SJIs)  and spectra of the Sun with high spatial resolution of  0\arcsec.16 pixel$^{-1}$ and temporal cadence as high as 1.5 s in four different wavelengths (\CII\ 1330, \SiIV\ 1400, \MgII\ k 2796, and \MgII\ wing 2830). For study of the network jets, we used \iris\ \SiIV\ slit-jaw images that have a temporal cadence of 21 seconds. Network jetlets show-up better in \SiIV\ slit-jaw images than in the other slit-jaw images. The FOV we investigate is not covered by the \iris\ slit.

Simultaneously, we used extreme-ultraviolet (EUV) images from AIA onboard \sdo\   to study the filament/filament channel. AIA captures high spatial resolution  (0\arcsec.6 pixel$^{-1}$)  full-Sun images, at 12 s temporal cadence, in seven different EUV channels. For our investigations, we used 304, 171, 193 and 211 \AA\ images and we found that the counter-streaming flows were best seen in 171 \AA\ images. 

To study the photospheric magnetic field evolution of the filament ends and jetlets, we used line of sight (LOS) magnetograms from HMI onboard \sdo. HMI LOS magnetograms have spatial resolution of 0\arcsec.5 pixel$^{-1}$, a temporal cadence of 45 s, and a noise level of about 7 G \citep{schou12,scherrer12,couvidat16}. We used \textit{GONG}  H$\alpha$ data to confirm that the filament was visible in H$\alpha$ filtergrams. 

 \iris, AIA and HMI data sets are co-aligned using SSW routines \citep{freeland98}. To enhance the visibility of weak minority-polarity flux patches at the edges the magnetic network lanes, we summed two consecutive magnetograms at each time step. 
 We have created movies, using \iris, AIA and HMI data, to follow the evolution of counter-streaming plasma flows and jetlets together with their underlying magnetic field. We have  overplotted contours, of $\pm$20 G, of  HMI LOS magnetograms. The white arrows and blues circles in the movies point to the jetlet spire and jetlet base, respectively. By westward or eastward flow, we mean plasma flows towards solar west or towards solar east, respectively.


\begin{figure}
	\centering
	\includegraphics[width=\linewidth]{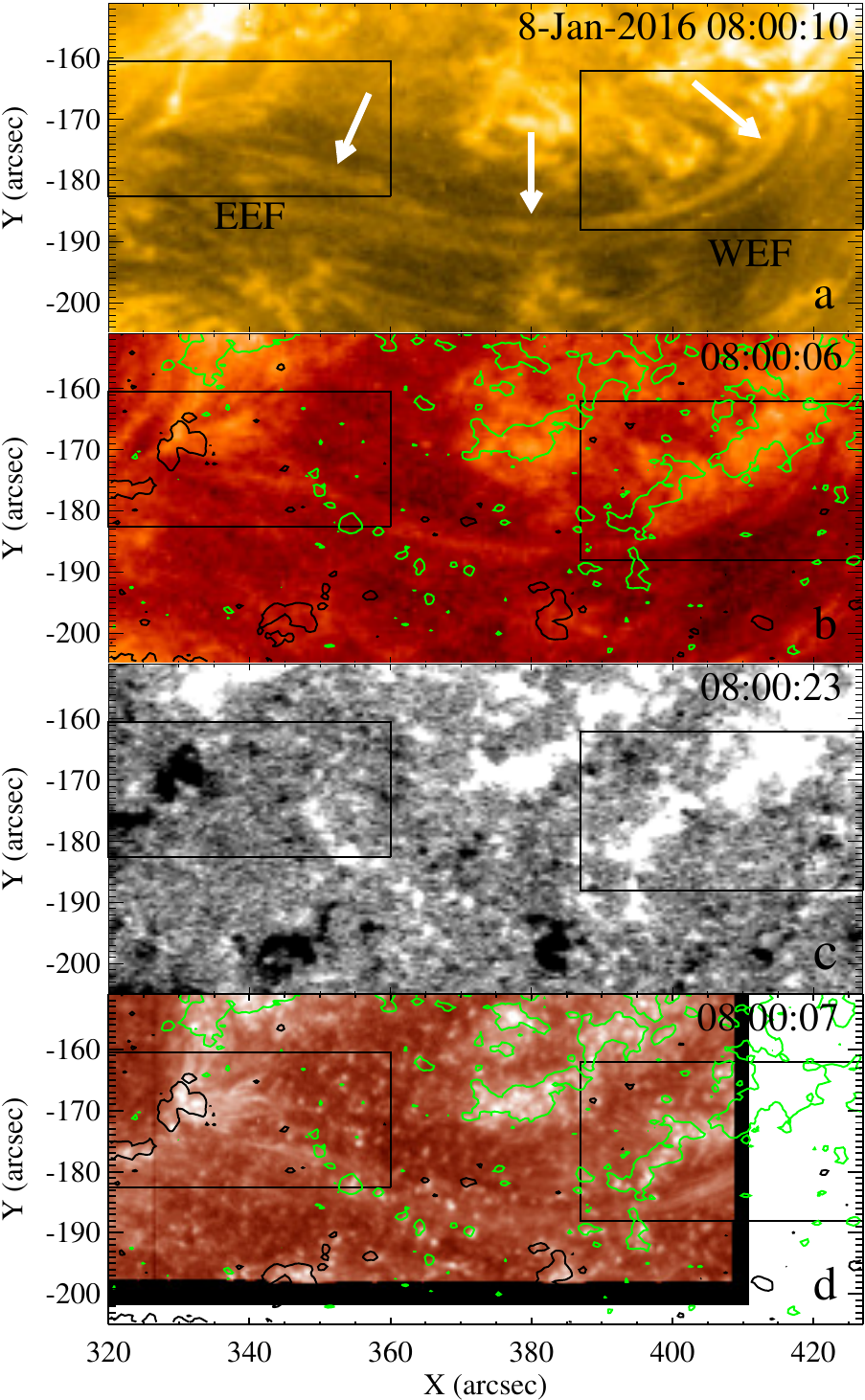}
	\caption{Overview of the filament observed on 8-Jan-2016. (a) AIA 171 \AA\ image of the filament. The white arrows point to the filament. (b) AIA 304 \AA\ image of the filament. (c) HMI magnetogram showing the photospheric  magnetic field of the same region. (d) \iris\ \SiIV\ SJI covering most of the same region. The black boxes show the locations of  network jetlets at the east end of the filament (EEF) and at the west end of the filament (WEF), and are the FOVs analyzed in detail in Figures \ref{fig2} and \ref{fig3}. The WEF is not covered by the \iris\ FOV  (the white region in d). In (b) and (d), $\pm$20 Gauss Green and black  contours (at 08:00:23UT)  outline the positive- and negative-polarity flux patches, respectively.	MOVIE1 is an animation of this Figure. } \label{fig1}
\end{figure} 

\section{Results}

\subsection{Overview}\label{over}

 Figures \ref{fig1h}a and \ref{fig1h}b show the filament observed by GONG in H$\alpha$ (see white arrows in Figure \ref{fig1h}a), and Figure \ref{fig1h}c shows the same filament in AIA 171 \AA\ image. The white contour outlines the filament/filament channel in the \Halpha\ image, and the FOV of the white box is used for detailed analysis of \sdo\ and \iris\ data. The filament has a wider appearance in AIA EUV images than in H$\alpha$ images, particularly in the middle segment. 
 	That the measured width of filaments is wider in EUV images than in H$\alpha$ images has been reported by \cite {hei01a,aulanier02,hei03,dudik08,parenti14,diercke18}.
 	
 The filament/filament channel in our study is similar to that reported by \cite{alexander13}.  Both in \cite{alexander13} and in our observations, the \Halpha\ filament (i) appears narrower than the filament/filament channel in AIA and Hi-C EUV images (Figure \ref{fig1h}; see Figure \ref{fig6} for our  comparison of measured widths), and (ii) does not show the two-strand structure  seen in EUV images that show the counter-streaming flows. On the basis of these strong qualitative similarities between our filament/filament channel and the filament/filament channel of \cite{alexander13}, we believe that we are observing the same phenomenon as observed by \cite{alexander13}, which they call ``counter-streaming" flows. We therefore follow the same terminology.
 Further, because the middle segment of the “filament” in the AIA EUV images is wider than the filament in the \Halpha\ image, we use the term filament/filament channel for the two-stranded EUV structure that closely tracks the \Halpha\ filament. 
 
 Figure \ref{fig1} shows the same H$\alpha$ filament observed with AIA and \iris\ images (see white arrows). It is an intermediate filament \citep{gai97,gai98} in the enhanced network remnant of decayed active region AR 12476, with  east foot anchored in negative polarity magnetic flux,  and the west foot rooted in positive polarity flux. The \Halpha\ filament resides above a weak magnetic neutral line,  the northern/southern part of which has dominant positive/negative magnetic polarity  (see, Figure \ref{fig1}c). 
 
 During  one hour of \iris\ coverage, we noticed several network jetlets at the east end of the filament (EEF), (see Figure \ref{fig1}d) that occur at the edges of the magnetic network lane between the majority-polarity negative magnetic flux of the network lane and a merging weak minority-polarity (positive) magnetic flux  patch. At the same time, these jetlets also appear in AIA images but not as clearly as in \iris\ SJIs (see MOVIE1). We use AIA 171 and 304 \AA\ images to study the jetlets at the west end of the filament (WEF) because the WEF  is not covered by the \iris\ FOV (Figure \ref{fig1}d).  The properties of these jetlets are similar to the jetlets observed by \cite{panesar18b,panesar19}. In Table \ref{tab:list}, we  list 10 randomly-selected network jetlets that occur at the two ends of the filament.
 Here, we present jetlets (2 and 5) from the east end of the filament/filament channel and jetlets (8 and 9) from the  west end of the filament/filament channel . 
 

\begin{figure*}
	\centering
	\includegraphics[width=0.95\linewidth]{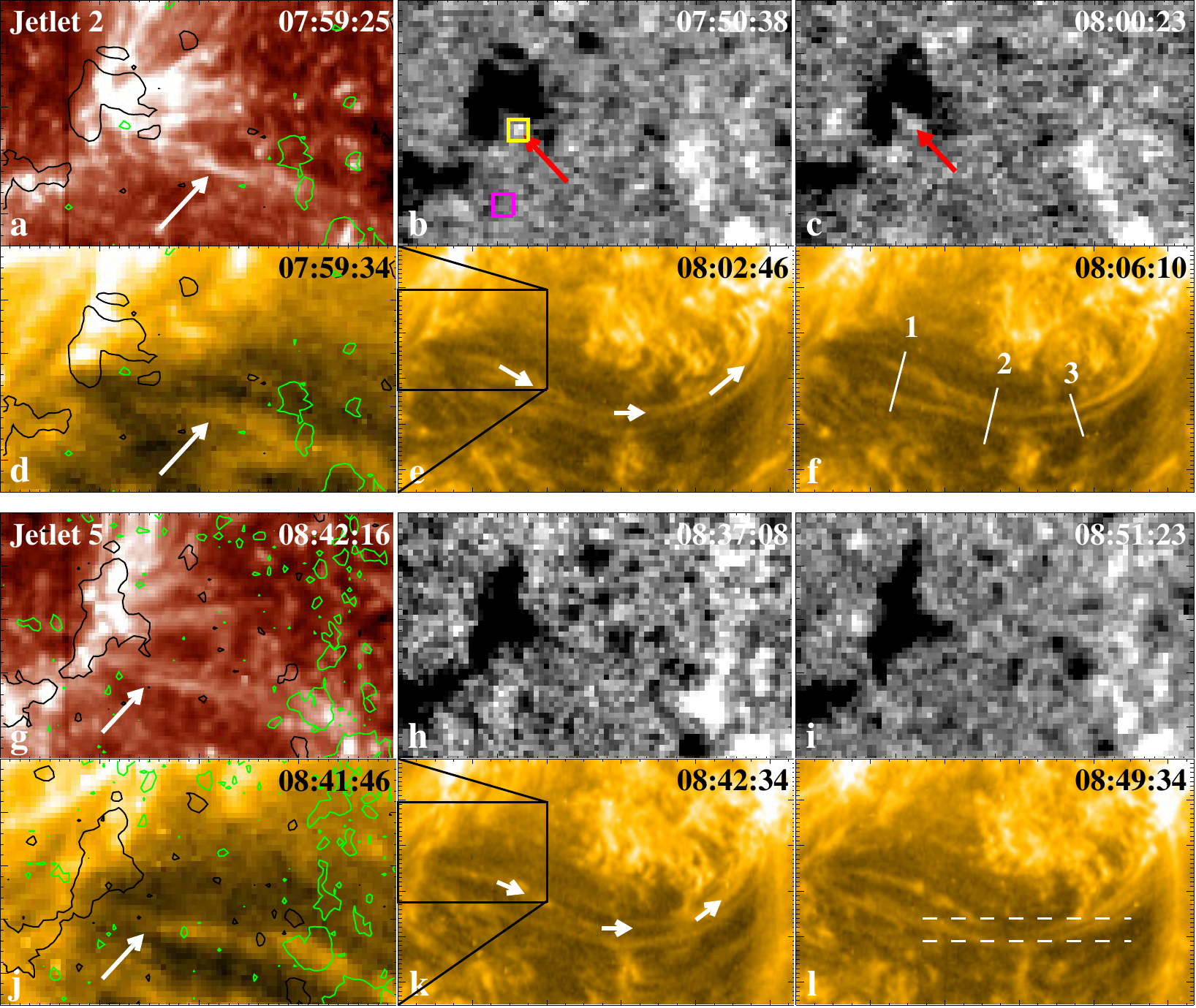}
	\caption{ Jetlets at the east end of the filament/filament channel. Panels (a) and (d) show the \iris\ \SiIV\ SJI and AIA 171 \AA\ images of the onset of jetlet 2; the white arrows point to the jetlet spire. Panels (b) and (c) show the HMI magnetograms of the same region; the red arrow in each magnetogram points to the minority-polarity canceling flux patch, the yellow box shows the area that is used to calculate the flux evolution plot shown in Figure \ref{fig5}a, and the magenta box shows the area that is used to estimate the noise in the measured flux plotted in Figure \ref{fig5}a. Panels (e) and (f) show  AIA 171 images of the filament and have the same FOV as in Figure \ref{fig1}. The white arrows show the direction of the westward flow that is driven by the jetlet shown in a and d; the black box shows the FOV displayed in panels a, b, c, and d, and the white lines (1-3) in (f) show the cuts across the filament threads used for measuring the intensity for the plot shown in Figure \ref{fig6}. In panels (a and d) HMI contours (of level $\pm$20 G) of 07:49:53 are overlaid, where  green and black represent positive and negative polarities, respectively. Panels (g) and (j) show the \iris\ \SiIV\ SJI and AIA 171 \AA\ images of  jetlet-5 from the same region. The black box in (k) shows the FOV displayed in panels (g, h, i, and j). The white dashed lines in panel (l)  show the two east–west cuts for the two time-distance maps in Figure \ref{fig4}. In panels (g and j) HMI contours (of level $\pm$20 G) of  08:42:23 are overlaid, where  green and black represent positive and negative polarities, respectively. Other annotated features are the same as above.
		Animation (MOVIE2) of this Figure is available.} \label{fig2}
\end{figure*} 

\subsection{Counter-Streaming Flows and Network Jetlets at the East End of the Filament/Filament channel }\label{left}


\textit{Jetlet-2:} Figures \ref{fig2}(a) and (d) show the example of jetlet-2, from Table \ref{tab:list}, observed by \iris\ and AIA. The accompanying movies (MOVIE1 and MOVIE2) show the complete evolution of the jetlet.
At 07:57:38, the jetlet spire starts to rise (MOVIE1 and MOVIE2) and it continues to grow until 08:00:49. We also see brightening in \SiIV\  SJIs at the base of the jetlet at about 07:58:21 (see blue circle at 07:59:25 in MOVIE2). At 07:58:21, a faint brightening at the base also appears in 171 \AA\ images (MOVIE2). Soon after the start of the jetlet spire,  plasma  begins to flow along the filament/filament channel strands (shown by the arrows in Figure \ref{fig2}e and MOVIE). The westward flow shows that the filament/filament channel  is made up of several horizontal  threads, like those in the filament studied by \cite{alexander13}.

The east end of the filament/filament channel  is rooted at the west edge of a negative-polarity-network patch of magnetic flux (Figures \ref{fig1}b,c). \iris\ and HMI movies show network jetlets that originate from that edge of the negative network flux patch. We observe an  opposite-polarity weak magnetic flux patch present near that edge of the negative-polarity network flux patch, and that the flow-driving jetlet (jetlet-2) erupted from the neutral line between the negative-polarity network flux patch and the weak positive flux patch (see red arrow in Figures \ref{fig2}b,c), which was merging into the negative network flux patch and evidently canceling with it. 

\begin{figure*}
	\centering
	\includegraphics[width=0.9\linewidth]{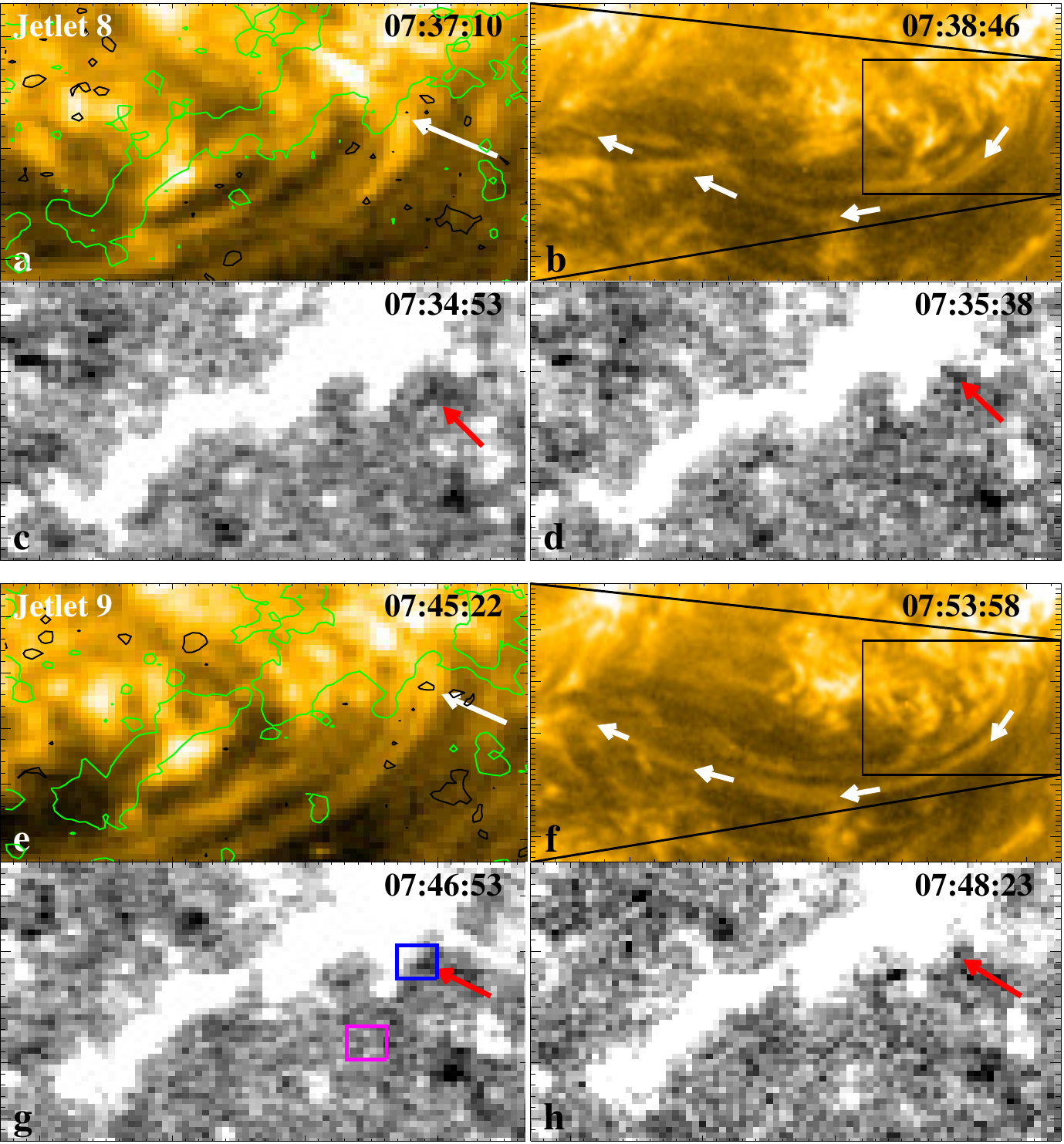}
	\caption{Jetlets at the west end of the filament/filament channel . Panel (a)  shows an  AIA 171 \AA\ image of the onset of jetlet-8. The white arrow in (a) points to the base of the jetlet spire. HMI contours (of level $\pm$20 G) of magnetogram of 07:35:38 are overlaid onto panel (a), where  green and black represent positive and negative polarities, respectively. Panels (c) and (d) show the HMI magnetogram of the same region; the red arrow points to the minority-polarity (negative) canceling flux patch. Panel (e)  shows an AIA 171 \AA\ image of  jetlet-9; HMI contours (of level $\pm$20 G) of magnetogram of 07:46:53 are overlaid onto panel (e), where  green and black represent positive and negative polarities, respectively.
		The white arrows in (b) and (f) show the path of the eastward flow. The black boxes in  (b,f) show the FOV displayed in (a, c,d, e, g, and h).  The blue box in (g) shows the area  used to calculate the flux plot shown in Figure \ref{fig5}b and the magenta box shows the area that is used to estimate the error bars.
		Animation (MOVIE3) of this Figure is available.} \label{fig3}
\end{figure*} 

\begin{figure*}
	\centering
	\includegraphics[width=\linewidth]{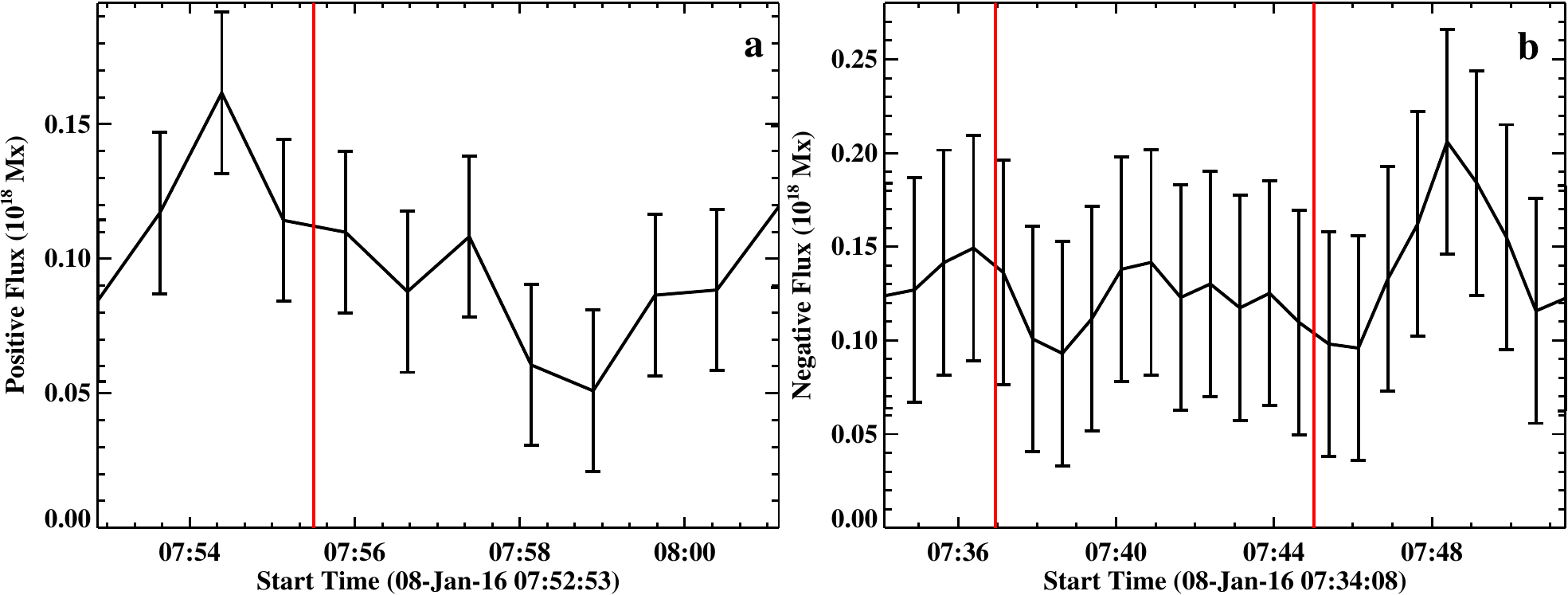}
	\caption{Magnetic flux plots for jetlets (2, 8, and 9) of Table \ref{tab:list}. Panel (a) shows the positive flux plot as a function of time computed inside the yellow box of Figure \ref{fig2}b.  Panel (b) shows the negative flux plot as a function of time computed inside the blue box of Figure \ref{fig3}g. The three red vertical lines mark the onsets of the three jetlet spires. Error bars in each panel represent the the uncertainty in the measured magnetic flux from magnetograph noise, as estimated in the way described in Section \ref{left}.} \label{fig5}
\end{figure*}

Figure \ref{fig5}a shows the minority-polarity (positive) flux plot of the region that is bounded by the yellow box of  Figure \ref{fig2}b. The positive flux decreases between 07:55 and 07:58 UT, presumably due to flux cancelation that prepares and triggers the eruption of jetlet-2 (see MOVIE2). During the same time, there appears to be also a decrease in negative flux at the adjacent edge of the negative-network flux patch (see red arrow in Figures \ref{fig2}b,c), and this suggests that flux cancelation is occurring. Presumably,  the flux cancelation prepares and triggers the eruption of the jetlet,  and that jetlet drives the westward streaming flow in the  filament/filament channel  strand. Large-scale coronal jet eruptions \citep{panesar16b,panesar17,sterling17,panesar18a,mcglasson19} and jetlets \citep{panesar18b,panesar19} are also seen to be prepared and triggered by flux cancelation at an underlying neutral line.

To estimate the uncertainty in the measured flux plotted in Figure \ref{fig5}a, we selected an apparently flux-free region (outlined by the magenta box in Figure \ref{fig2}b, which is of the same size as the yellow box in which the flux for the plot in Figure \ref{fig5}a was measured) that has no visibly discernible non-noise magnetic flux in it throughout the time period of the flux plot in Figure \ref{fig5}a. 
The uncertainty (from magnetogram noise) in the flux measured in the yellow box was estimated by smoothing (by a factor of two) and adding up the total unsigned flux in all of the pixels in the magenta box in each frame.  We then averaged the values of total unsigned flux over time. The half of this value is 3.6 $\times$ 10$^{16}$ Mx, which is plotted in Figure \ref{fig5}a as error bars for positive magnetic flux. The error bar for Figure \ref{fig5}b and the uncertainty in the Figure \ref{fig3} magnetogram field strength were estimated in the same way, using the magenta box in Figure \ref{fig3}g: we obtained $\pm$  6.4 $\times$ 10$^{16}$ Mx for the error bars in Figure \ref{fig5}b for negative magnetic flux. Note that the measured flux (the solid curve) in Figure \ref{fig5}a and in Figure \ref{fig5}b is above zero by more than the downward extents of the error bars for the entire time interval of each of the two plots; this demonstrates that the flux in each of the two measured minority flux patches is definitely above noise, and hence that each of these minority flux patches is real, not just noise.

\textit{Jetlet-5:}  Figures \ref{fig2}(g)-(j), and accompanying movies (MOVIE1 and MOVIE2), show the example of jetlet-5 from Table \ref{tab:list}. 
The white arrow points to the jetlet spire, which was visible between 08:41:54 and 08:43:19 in \SiIV\ SJIs (MOVIE1 and MOVIE2), as well as in  AIA images. \iris\ images also show base brightening during the onset of the jetlet, but the AIA images do not show any corresponding brightening at the  base of the jetlet. In MOVIE1, we can clearly see that westward flow in the filament originate from this jetlet (the direction of the flows in shown in Figure \ref{fig2}k). 

Jetlet-5 is rooted at the edge of a negative-polarity network magnetic flux lane (Figures \ref{fig2}(g) and (h)). Unlike the case of jetlet-2, we do not see any visible flux cancelation before and during the  onset of this jetlet. But we do see a decrease in flux at the edge of the negative-polarity network flux lane from where the jetlet starts from before to after the jet erupts. Based on our previous experience, we suspect that there was some weak small-scale minority-polarity flux present at the base of the jetlet below the noise level of HMI but we cannot rule out other possibilities with the given data set

Similarly, there are other jetlets that come from the same network flux lane (see Table \ref{tab:list}). MOVIE1 shows the complete evolution of all the jetlets and westward-streaming flows in the filament that come from  the jetlets. Jetlets (2-6) show base brightening during the eruption onset; these base brightenings are clearly visible in the \iris\ images. All the jetlets are rooted at the edge of the negative magnetic network flux lane. Right after a jetlet starts, usually flow along a filament/filament channel  thread stemming from the jetlet begins to move westward along the thread. Figure \ref{fig4}a shows a running-difference time-distance map along the northern dashed line of Figure \ref{fig2}l. It shows obvious westward flows in the filament's threads; some of the flows are highlighted with red-dashed arrows (see Section \ref{flow}). These examples clearly show that the flows are apparently driven by  network jetlet eruptions at sites of magnetic flux cancelation.

\subsection{Counter-Streaming Flows and Network Jetlets at the West End of the Filament/Filament channel } \label{right}

\textit{Jetlet-8:} Figure \ref{fig3}(a) shows  jetlet-8 in an AIA 171 \AA\ images. Note that the west end of the filament/filament channel  is not covered by \iris. The accompanying movies (MOVIE1 and MOVIE3) show the complete evolution of  jetlet-8. AIA 171 \AA\ images show  faint base brightening in jetlet-8 (at 07:32:34, MOVIE3). The jetlet spire starts to rise at about 07:34:58 and continues to grow (white arrow in MOVIE3).  Shortly after the onset of the jetlet spire, plasma starts to move eastward along a filament/filament channel  thread stemming from the site of the jetlet (see white arrows in Figure \ref{fig3}b and MOVIE1).  

The jetlet erupts from the east edge of the majority-polarity (positive) network flux lane (Figures \ref{fig3}a,c). The red arrows in Figures \ref{fig3}c,d point to the weak minority-polarity negative flux patch that converges towards the positive flux patch and plausibly prepares and triggers the jetlet eruption. Figure \ref{fig5}b shows a plot of the negative flux patch versus time during the onset time of  jetlet-8 (the first red line in Figure \ref{fig5}b).  Although, the MOVIE3 shows flux cancelation more clearly, as mentioned before, the error bars in Figure \ref{fig5}b are large, suggesting only marginal flux cancelation at the base of the jetlet.

\textit{Jetlet-9:} In  Figure \ref{fig3}(e), we show an AIA 171\AA\ image of jetlet-9. The detailed evolution of the jetlet and its photospheric magnetic field evolution can be seen in accompanying movies (MOVIE1 and MOVIE3). A faint jetlet spire extends outward at 07:45:22 (shown with an arrow in Figure \ref{fig3}e and MOVIE3).  Immediately after the appearance of the spire, eastward flow begins to move in the filament's thread, along a thread stemming from the site of the jetlet. The white arrows in Figure \ref{fig3}f highlight  the eastward flow in the thread of the filament/filament channel . Unlike for jetlet-8, we do not see any brightening at the base of jetlet-9. 

 Figures \ref{fig3}(g,h) display the photospheric magnetic field evolution of the base of the jetlet. The jetlet is rooted at the neutral line between a majority-polarity (positive) network flux lane and minority-polarity (negative) weak flux patch (red arrows in Figures \ref{fig3}(g,h)). Figure \ref{fig5}b shows the minority-polarity flux patch versus time. It shows a magnetic flux evolution during the  during the eruption of jetlet-9 (second red line).
 
 It is clear that the eastward-streaming flows from the west end of the filament/filament channel  are also driven by jetlets, similar to the westward flows from  the east end of the filament. The timings of significant jetlets  are listed in Table \ref{tab:list}. Just after each of these jetlets, plasma  starts to flow eastward along the filament/filament channel thread. Figure \ref{fig4}b displays  clear evidence of the eastward flows in the southern thread of the filament. Some of the stronger flows are highlighted with red dashed arrows.

\subsection{Flow Speed}\label{flow}

\begin{figure}
	\centering
	\includegraphics[width=\linewidth]{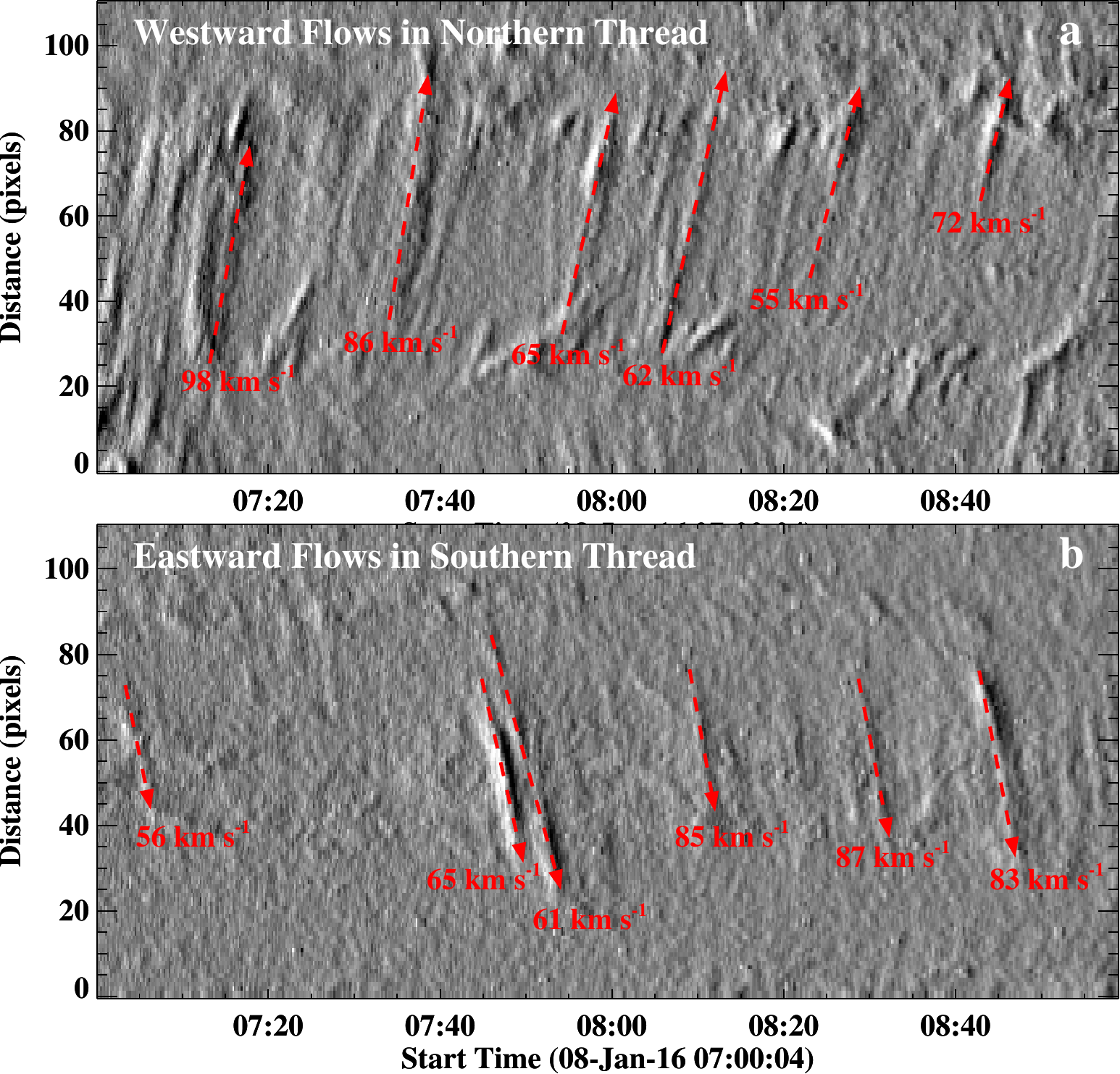}
	\caption{Counter-streaming flows in  two filament/filament channel  threads. Panel (a) and Panel (b) respectively show the AIA 171 \AA\ running-difference time-distance images along the northern and southern dashed lines drawn in Figure \ref{fig2}l. They show  the flows along the filament's northern thread and the southern thread, respectively. The red arrows indicate the  flow direction in some of the stronger flows.  } \label{fig4}
\end{figure}

Figure \ref{fig4} shows  horizontal flows in the northern thread and in the southern thread of the filament/filament channel. To present a clear picture of  counter-streaming flows, we made a running-difference time-distance map from along each of the two white dashed lines of Figure \ref{fig2}l.
These two maps  clearly show the plasma flows in opposite directions along the filament/filament channel  threads (see red arrows). In the northern thread the plasma travels towards  solar west whereas in the southern thread the plasma travels towards  solar east. Thus, these are counter-streaming flows. 

One can clearly see the persistent and continuous counter-streaming flows in the two strands from 07:54 to 08:15 UT (in Movie1) in 171\AA\ images. The observed counter-streaming flows are similar to the flows that are reported by \cite{alexander13} in two adjacent strands of a filament in Hi-C 193 \AA\ images. The counter-streaming flows are sporadic at times, particularly in the beginning of the movie (Movie 1), in the manner of the counter-streaming flows reported by \cite{diercke18}.

We also observed  additional flows in the filament/filament channel  that are not on the path of either of these two selected time-distance cuts. For example in Figure \ref{fig4}b, some strong eastward flows that are driven by jetlets (7,8) (between 07:15 and 07:40 UT) and are not discernible due to the selection of this time-distance cut.

Some of the counter-streaming flows are highlighted with red-dashed arrows and their speeds are noted in 
Figure \ref{fig4}. The westward and eastward flows move with an average speed of 73$\pm$16 \kms\  and 73$\pm$13 \kms, respectively. Note that our measured speeds are lower limits on the full speeds of the 3D flows because only the projected (proper-motion) speed is measured; the full speed is the root-mean-square of proper-motion speed (which we measure) and the speed along the line of sight (which speed we do not measure, but we expect it to be small compared to the proper-motion speed in our case because the filament is on the central disk and the field in it is roughly horizontal). The observed flow speeds are in agreement with the speeds (70 \kms) obtained by \cite{alexander13} for an active-region filament, using Hi-C data, whereas \cite{zirker94} and \cite{zir98} reported flow speeds in the range of 5-20 \kms\ for quiet-region filaments, using H$\alpha$ data.

\subsection{Thread Width}
\begin{figure*}
	\centering
	\includegraphics[width=0.495\linewidth]{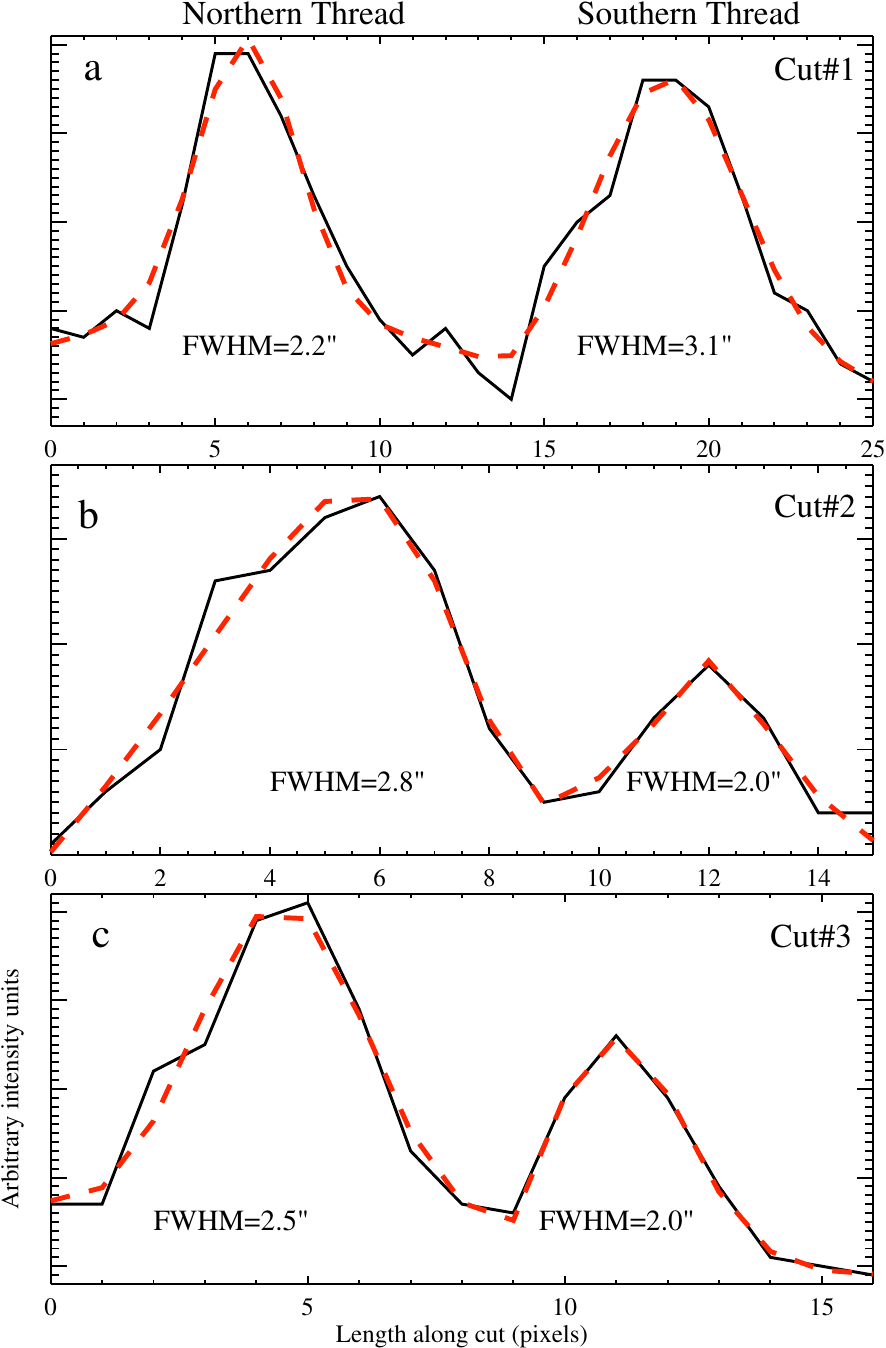}
	\put(-175,400){Width of the AIA threads}
	\includegraphics[width=0.49\linewidth]{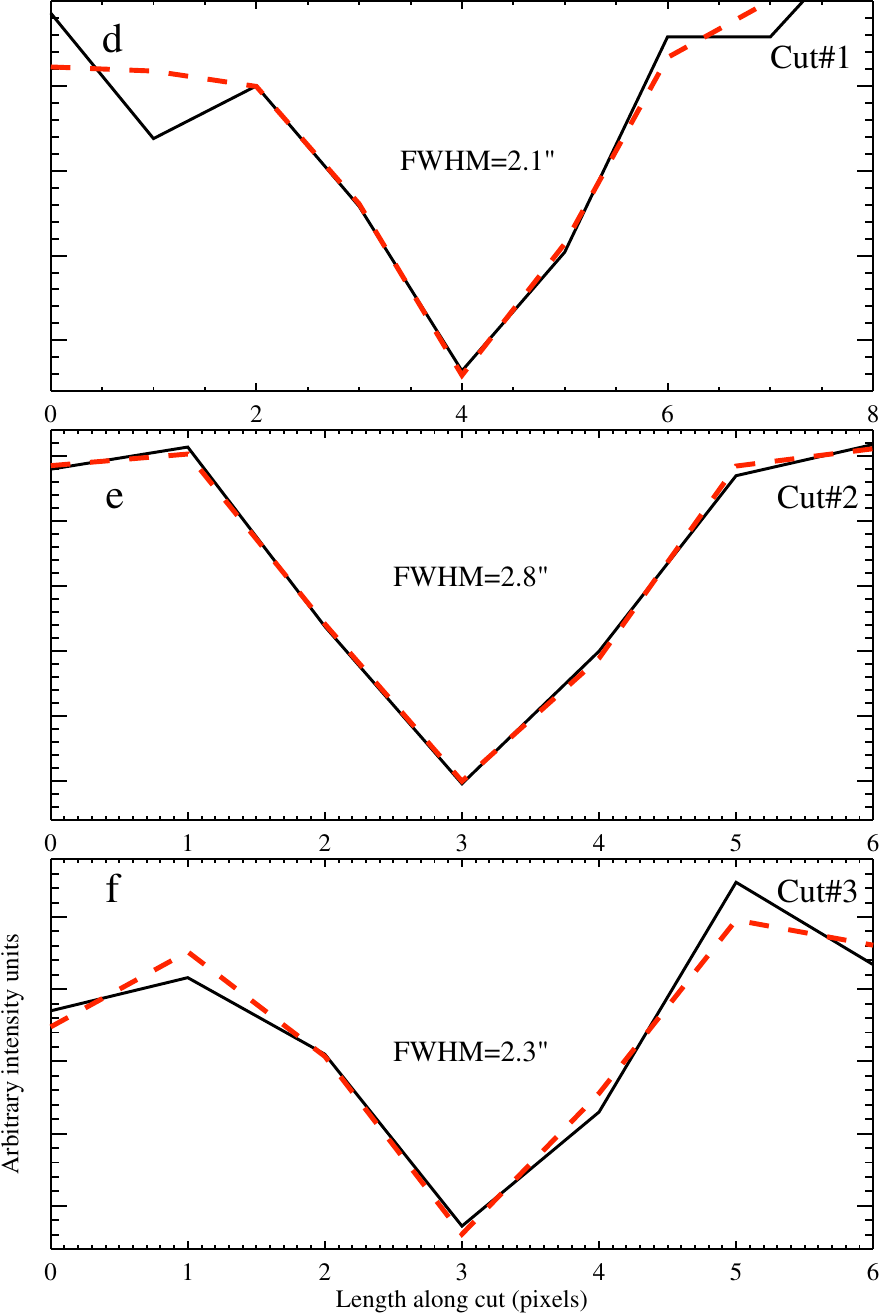}
	\put(-175,400){Width of the \Halpha\ threads}
	\caption{Widths of the filament/filament channel in EUV/AIA and \Halpha\ images. Panels (a), (b), and (c)  show the AIA 171 \AA\ intensity curve (black) and the fitted Gaussian function (red) along the cuts (1--3) of Figure \ref{fig2}f, respectively. Panels (d), (e), and (f)  show the \Halpha\ intensity curve (black) and the fitted Gaussian function (red) along the same cuts in the \Halpha\ image in Figure \ref{fig1h}. Note that the range on x-axes is not the same in all panels. Pixel size of AIA and GONG images are 0.6\arcsec\ and 1\arcsec, respectively.}  \label{fig6}
\end{figure*}

We have measured the widths of the filament/filament channel (using AIA 171 \AA\ and \Halpha\ images in Figure \ref{fig1h}), from the intensity-distance plots from transverse cuts at three places along the filament/filament channel (Figures \ref{fig6}). In the plots from the AIA image, the first peak (Figures \ref{fig6}a-c) corresponds to the northern thread of the filament/filament channel and the second peak corresponds to the southern thread of the filament/filament channel. There is a decrease in  intensity in-between the two threads in the gap between the two threads. Similarly, Hi-C observations \citep{alexander13} also show a gap between adjacent  threads. According to \cite{zir98}, counter-streaming in filament threads might be either superimposed along the line-of-sight on the same thread or there is a separation between  filament threads having opposite flows. Figures \ref{fig6}d-f show the measured widths of the filament in the \Halpha\ image.  These plots confirm the visual impression from Figure \ref{fig1h} that in the middle of the filament/filament channel (well away from each end), the \Halpha\ filament is narrower than the AIA EUV filament/filament channel.

The average  widths (using AIA images) of the northern and southern threads are 3.7 $\pm$ 3\arcsec\ and 4.0 $\pm$ 2\arcsec, respectively. The observed AIA thread widths are wider than the widths (0.8\arcsec) obtained by \cite{alexander13},  using high-resolution 193 \AA\ images from  Hi-C,  and  the widths (0.3\arcsec) observed by \cite{lin05b}, using high-resolution H$\alpha$ images from the Swedish Solar Telescope. Apparently the filament threads seen in our AIA 171 \AA\ images are bundles of a few finer threads of narrower widths such as seen in higher-resolution \Halpha\ images of other filaments.

\section{Discussion}

We have investigated the driving mechanism of counter-streaming flows in a solar filament/filament channel  using \sdo\ and \iris\ data. We show the first direct observations of counter-streaming flows in a filament/filament channel  from AIA, without using any image enhancement techniques. We find that (i)  opposite ends of the filament/filament channel  are  at opposite-polarity magnetic network flux lanes; (ii) repeated small-scale jets (jetlets) erupt from  the flux-lane edges  where the ends of the filament/filament channel  are rooted; (iii) the jetlet eruption at the two ends drive the observed  counter-streaming flows in the  filament/filament channel; (iv) some of the jetlets appear to erupt from sites of  flux cancelation between  majority-polarity network flux and merging minority-polarity flux, reminiscent of coronal jets; and (v) \iris\ \SiIV\ SJIs show brightening at the base of the jetlets; this brightening is similar to the base-brightening in coronal jets. 

According to  \cite{chen14} longitudinal oscillations (when they are not in phase) in filament's threads in \Halpha\ observations could appear as bi-directional flows. Whereas \cite{low05} suggested that bi-directional flows occur due to non-equilibrium in magnetic forces at the footpoints of  filament threads. \cite{shen15} argued that counter-streaming flows are possibly caused by  magnetic reconnection at a separator between the prominence bubble and the overlying-magnetic-field dips. \cite{wang18} present high-resolution \Halpha\ observations of counter-streaming flows that are obviously flows of plasma along a filament’s fibrils.

Our observations clearly show that plasma flows in the filament/filament channel  threads are driven by  network jets. The AIA 171 \AA\ movie shows clear evidence of continuous counter-streaming flows  throughout the filament structure (they are not localized to some places), and that the flows originate from the two ends of the filament.  The counter-streaming flows travel along  two different strands, and move with a speed of about 70 \kms\ in both  threads. These flows  are similar to the counter-streaming flows observed by Hi-C \citep{alexander13} where the authors suggested that the oppositely directed flows were in  separate strands (threads) of the same filament, and they observed  flow speeds in both threads similar to those in the present observations. The AIA filament/filament channel  threads that we observe are wider than those observed by \cite{lin05b} and by \cite{alexander13}, perhaps mostly due to AIA's lower spatial resolution.  The average width of our filament/filament channel  threads is 4.0\arcsec, whereas \cite{lin05b} obtained 
0.3\arcsec\ using SST H$\alpha$ images and \cite{alexander13} reported 0.8\arcsec\ using Hi-C data. Thus, our observations provide information on counter-streaming flows in  wider filament threads (or perhaps in bundles of threads) than found from those higher-resolution observations.

The cool plasma moves with an average speed of about 70 \kms\ in  both threads. The observed EUV flows have higher speeds than the flows observed  using H$\alpha$ data in quiet region filaments. For example, \cite{zir98} and \cite{lin03} obtained plasma flows speeds between 5-20 \kms\ and 8 \kms, respectively, using H$\alpha$ observations of quiet-region filaments. Our measured flow speeds are in agreement with the EUV observations reported by \cite{lab10}  and \cite{alexander13}, who observed velocities of 70 \kms\ along  individual threads of active-region filaments, similar to that seen in our observations of an intermediate filament.

Network jetlets tend to erupt repeatedly at the edges of strong magnetic network flux clumps are at the boundaries of supergranule convection cells \citep{tian14,panesar18b,panesar19}. Jetlets (1, 2, 3, 5, and 6) occur at the sites of discernible flux cancelation, and base brightening appears at the canceling neutral line. 
In jetlet 4  we do see a signature of minority-polarity flux at the base of the jetlet but do not observe  clear evidence of flux cancelation. In this jetlets, we do observe brightening at the base.  It is possible that a weak minority-polarity flux is present there at the edge of the network flux, but not detected by HMI. 
Since \iris\ FOV does not cover the west end of the filament, there we use only AIA 171\AA\ images to study the jetlets and their base brightenings. Jetlets (7 and 8) each show  base brightening and each brightening occurs at a cancelation neutral line, similar to the \iris\ jetlets observed at the east end of the filament. These results are apparently consistent with flux cancelation at the neutral line (between  majority-polarity network flux and a  minority-polarity flux patch) preparing and triggering  multiple jetlet eruptions \citep{panesar18b,panesar19}. The base brightening that occurs at a canceling neutral line and/or at the edge of network  flux is possibly a miniature version of the jet-base-brightening that  occurs at a canceling neutral line before and during coronal jets \citep{panesar16b,panesar17,panesar20}. The discernible flux cancelation plausibly prepares and triggers the multiple jetlets at the both ends of the filament/filament channel  that drive continually repeated counter-streaming  flows in the filament/filament channel.

If these jetlets are miniature versions of coronal jets then the base brightenings would result from the \textit{internal reconnection} (it occurs within the legs of erupting minifilament flux rope and its enveloping field; \citealt{sterling15}), and the spire would result from the \textit{external/interchange reconnection} (when the erupting flux-rope envelope and flux rope reconnect with encountered far-reaching magnetic field). In the present case, the jetlets shoot out horizontally along the horizontal magnetic field (in the filament/filament channel). We infer that the plasma flows are injected into the filament threads via the external reconnection that makes the jetlet spire. The counter-streaming flows may be further magnetically driven by magnetic-twist waves from the external reconnection of the erupting flux rope if that flux rope contains twist \citep{moore15}.

Jets and jetlets are also seen to occur at other sites of  flux cancelation (sometimes accompanied, or preceded, by flux emergence and/or flux coalesce) in  mixed polarity photospheric flux at the base of  far-reaching coronal loops \citep[e.g.][]{tiwari19,panesar19} and they might contribute to  heating of  coronal loops. Upflows in polar crown prominences have been observed driving flux cancelation at the feet of prominence pillars/barbs \citep{panesar14c}.
Similarly, our network jetlets (evidently prepared and  triggered by flux cancelation) at both feet of the filament/filament channel  are driving  the flows in the filament's threads. As in filament threads,  blueshifts and redshifts in coronal loops might be due to explosive events at the base \citep[e.g.][]{srivastava2020}. Whether such flows in coronal loops come from jetlet eruptions remains to be established.

Our \sdo\ and \iris\ observations show  examples of recurrent counter-streaming flows along  filament/filament channel  threads, and that these counter-streaming flows are clearly driven by  network jetlet eruptions that occur at  both ends of the filament/filament channel. The network jetlets erupt from the edges of magnetic network flux lanes and therefore plausibly are small-scale
versions of the larger jetlets and coronal jets, most or all of which are prepared and triggered to erupt by flux cancelation. 

	There have been several numerical models of larger-scale jets  (e.g., \citealt{cheung15}, \citealt{wyper18}, and references therein), but formation of small scale jets, i.e., network jets/jetlets, have not been well simulated. It will be interesting to see whether in a suitable magnetic configuration such counter-streaming flows as reported here are produced by simulated jetlets erupting from the ends of a filament field.

\acknowledgments
We thank an anonymous referee for constructive comments. NKP acknowledges support from NASA’s SDO/AIA (NNG04EA00C) and HGI grant  (80NSSC20K0720).  SKT gratefully acknowledges support by NASA contracts NNG09FA40C (IRIS), and NNM07AA01C (Hinode). RLM and ACS acknowledge the support from the NASA HGI program.
We acknowledge the use of  \iris,  \sdo/AIA,  \sdo/HMI, and GONG data. AIA is an instrument onboard the Solar Dynamics Observatory, a mission for NASA’s Living With a Star program.  \iris\ is a NASA small explorer mission developed and operated by LMSAL with mission operations executed at NASA Ames Research center and major contributions to downlink communications funded by ESA and the Norwegian Space Centre.  This work has made use of NASA ADSABS and Solar Software.

\bibliographystyle{aasjournal}

\begin{thebibliography}{}
	\expandafter\ifx\csname natexlab\endcsname\relax\def\natexlab#1{#1}\fi
	
	\bibitem[{{Ahn} {et~al.}(2010){Ahn}, {Chae}, {Cao}, \& {Goode}}]{ahn10}
	{Ahn}, K., {Chae}, J., {Cao}, W., \& {Goode}, P.~R. 2010, \apj, 721, 74
	
	\bibitem[{{Alexander} {et~al.}(2013){Alexander}, {Walsh}, {R{\'e}gnier},
		{Cirtain}, {Winebarger}, {Golub}, {Kobayashi}, {Platt}, {Mitchell},
		{Korreck}, {DePontieu}, {DeForest}, {Weber}, {Title}, \&
		{Kuzin}}]{alexander13}
	{Alexander}, C.~E., {Walsh}, R.~W., {R{\'e}gnier}, S., {et~al.} 2013, \apjl,
	775, L32
	
	\bibitem[{{Aulanier} \& {Schmieder}(2002)}]{aulanier02}
	{Aulanier}, G., \& {Schmieder}, B. 2002, \aap, 386, 1106
	
	\bibitem[{{Babcock} \& {Babcock}(1955)}]{bab55}
	{Babcock}, H.~W., \& {Babcock}, H.~D. 1955, \apj, 121, 349
	
	\bibitem[{{Chae} {et~al.}(2007){Chae}, {Park}, \& {Park}}]{chae07}
	{Chae}, J., {Park}, H.-M., \& {Park}, Y.-D. 2007, Journal of Korean
	Astronomical Society, 40, 67
	
	\bibitem[{{Chen} {et~al.}(2014){Chen}, {Harra}, \& {Fang}}]{chen14}
	{Chen}, P.~F., {Harra}, L.~K., \& {Fang}, C. 2014, \apj, 784, 50
	
	\bibitem[{{Cheung} {et~al.}(2015){Cheung}, {De Pontieu}, {Tarbell}, {Fu},
		{Tian}, {Testa}, {Reeves}, {Mart{\'\i}nez-Sykora}, {Boerner}, {W{\"u}lser},
		{Lemen}, {Title}, {Hurlburt}, {Kleint}, {Kankelborg}, {Jaeggli}, {Golub},
		{McKillop}, {Saar}, {Carlsson}, \& {Hansteen}}]{cheung15}
	{Cheung}, M. C.~M., {De Pontieu}, B., {Tarbell}, T.~D., {et~al.} 2015, \apj,
	801, 83
	
	\bibitem[{{Couvidat} {et~al.}(2016){Couvidat}, {Schou}, {Hoeksema}, {Bogart},
		{Bush}, {Duvall}, {Liu}, {Norton}, \& {Scherrer}}]{couvidat16}
	{Couvidat}, S., {Schou}, J., {Hoeksema}, J.~T., {et~al.} 2016, \solphys, 291,
	1887
	
	\bibitem[{{De Pontieu} {et~al.}(2014){De Pontieu}, {Title}, {Lemen}, {Kushner},
		{Akin}, {Allard}, {Berger}, {Boerner}, {Cheung}, {Chou}, {Drake}, {Duncan},
		{Freeland}, {Heyman}, {Hoffman}, {Hurlburt}, {Lindgren}, {Mathur}, {Rehse},
		{Sabolish}, {Seguin}, {Schrijver}, {Tarbell}, {W{\"u}lser}, {Wolfson},
		{Yanari}, {Mudge}, {Nguyen-Phuc}, {Timmons}, {van Bezooijen}, {Weingrod},
		{Brookner}, {Butcher}, {Dougherty}, {Eder}, {Knagenhjelm}, {Larsen},
		{Mansir}, {Phan}, {Boyle}, {Cheimets}, {DeLuca}, {Golub}, {Gates}, {Hertz},
		{McKillop}, {Park}, {Perry}, {Podgorski}, {Reeves}, {Saar}, {Testa}, {Tian},
		{Weber}, {Dunn}, {Eccles}, {Jaeggli}, {Kankelborg}, {Mashburn}, {Pust},
		{Springer}, {Carvalho}, {Kleint}, {Marmie}, {Mazmanian}, {Pereira}, {Sawyer},
		{Strong}, {Worden}, {Carlsson}, {Hansteen}, {Leenaarts}, {Wiesmann},
		{Aloise}, {Chu}, {Bush}, {Scherrer}, {Brekke}, {Martinez-Sykora}, {Lites},
		{McIntosh}, {Uitenbroek}, {Okamoto}, {Gummin}, {Auker}, {Jerram}, {Pool}, \&
		{Waltham}}]{pontieu14}
	{De Pontieu}, B., {Title}, A.~M., {Lemen}, J.~R., {et~al.} 2014, \solphys, 289,
	2733
	
	\bibitem[{{Diercke} {et~al.}(2018){Diercke}, {Kuckein}, {Verma}, \&
		{Denker}}]{diercke18}
	{Diercke}, A., {Kuckein}, C., {Verma}, M., \& {Denker}, C. 2018, \aap, 611, A64
	
	\bibitem[{{Dud{\'\i}k} {et~al.}(2008){Dud{\'\i}k}, {Aulanier}, {Schmieder},
		{Bommier}, \& {Roudier}}]{dudik08}
	{Dud{\'\i}k}, J., {Aulanier}, G., {Schmieder}, B., {Bommier}, V., \& {Roudier},
	T. 2008, \solphys, 248, 29
	
	\bibitem[{{Engvold}(1976)}]{eng76}
	{Engvold}, O. 1976, \solphys, 49, 283
	
	\bibitem[{{Engvold} {et~al.}(2001){Engvold}, {Jakobsson}, {Tand berg-Hanssen},
		{Gurman}, \& {Moses}}]{engvold01}
	{Engvold}, O., {Jakobsson}, H., {Tand berg-Hanssen}, E., {Gurman}, J.~B., \&
	{Moses}, D. 2001, \solphys, 202, 293
	
	\bibitem[{{Freeland} \& {Handy}(1998)}]{freeland98}
	{Freeland}, S.~L., \& {Handy}, B.~N. 1998, \solphys, 182, 497
	
	\bibitem[{{Gaizauskas}(1998)}]{gai98}
	{Gaizauskas}, V. 1998, in Astronomical Society of the Pacific Conference
	Series, Vol. 150, IAU Colloq. 167: New Perspectives on Solar Prominences, ed.
	D.~F. {Webb}, B.~{Schmieder}, \& D.~M. {Rust}, 257
	
	\bibitem[{{Gaizauskas} {et~al.}(1997){Gaizauskas}, {Zirker}, {Sweetland}, \&
		{Kovacs}}]{gai97}
	{Gaizauskas}, V., {Zirker}, J.~B., {Sweetland}, C., \& {Kovacs}, A. 1997, \apj,
	479, 448
	
	\bibitem[{{Harvey} {et~al.}(1996){Harvey}, {Hill}, {Hubbard}, {Kennedy},
		{Leibacher}, {Pintar}, {Gilman}, {Noyes}, {Title}, {Toomre}, {Ulrich},
		{Bhatnagar}, {Kennewell}, {Marquette}, {Patron}, {Saa}, \&
		{Yasukawa}}]{harvey96}
	{Harvey}, J.~W., {Hill}, F., {Hubbard}, R.~P., {et~al.} 1996, Science, 272,
	1284
	
	\bibitem[{{Heinzel} {et~al.}(2003){Heinzel}, {Anzer}, \& {Schmieder}}]{hei03}
	{Heinzel}, P., {Anzer}, U., \& {Schmieder}, B. 2003, \solphys, 216, 159
	
	\bibitem[{{Heinzel} {et~al.}(2001){Heinzel}, {Schmieder}, \&
		{Tziotziou}}]{hei01a}
	{Heinzel}, P., {Schmieder}, B., \& {Tziotziou}, K. 2001, \apjl, 561, L223
	
	\bibitem[{{Hirayama}(1985)}]{hir85}
	{Hirayama}, T. 1985, \solphys, 100, 415
	
	\bibitem[{{Labrosse} {et~al.}(2010){Labrosse}, {Heinzel}, {Vial}, {Kucera},
		{Parenti}, {Gun{\'a}r}, {Schmieder}, \& {Kilper}}]{lab10}
	{Labrosse}, N., {Heinzel}, P., {Vial}, J.-C., {et~al.} 2010, \ssr, 151, 243
	
	\bibitem[{{Lemen} {et~al.}(2012){Lemen}, {Title}, {Akin}, {Boerner}, {Chou},
		{Drake}, {Duncan}, {Edwards}, {Friedlaender}, {Heyman}, {Hurlburt}, {Katz},
		{Kushner}, {Levay}, {Lindgren}, {Mathur}, {McFeaters}, {Mitchell}, {Rehse},
		{Schrijver}, {Springer}, {Stern}, {Tarbell}, {Wuelser}, {Wolfson}, {Yanari},
		{Bookbinder}, {Cheimets}, {Caldwell}, {Deluca}, {Gates}, {Golub}, {Park},
		{Podgorski}, {Bush}, {Scherrer}, {Gummin}, {Smith}, {Auker}, {Jerram},
		{Pool}, {Soufli}, {Windt}, {Beardsley}, {Clapp}, {Lang}, \&
		{Waltham}}]{lem12}
	{Lemen}, J.~R., {Title}, A.~M., {Akin}, D.~J., {et~al.} 2012, \solphys, 275, 17
	
	\bibitem[{{Lin} {et~al.}(2005){Lin}, {Engvold}, {Rouppe van der Voort}, {Wiik},
		\& {Berger}}]{lin05b}
	{Lin}, Y., {Engvold}, O., {Rouppe van der Voort}, L., {Wiik}, J.~E., \&
	{Berger}, T.~E. 2005, \solphys, 226, 239
	
	\bibitem[{{Lin} {et~al.}(2003){Lin}, {Engvold}, \& {Wiik}}]{lin03}
	{Lin}, Y., {Engvold}, O.~R., \& {Wiik}, J.~E. 2003, \solphys, 216, 109
	
	\bibitem[{{Low} \& {Petrie}(2005)}]{low05}
	{Low}, B.~C., \& {Petrie}, G.~J.~D. 2005, \apj, 626, 551
	
	\bibitem[{{Mackay} {et~al.}(2010){Mackay}, {Karpen}, {Ballester}, {Schmieder},
		\& {Aulanier}}]{mackay10}
	{Mackay}, D.~H., {Karpen}, J.~T., {Ballester}, J.~L., {Schmieder}, B., \&
	{Aulanier}, G. 2010, \ssr, 151, 333
	
	\bibitem[{{Martens} \& {Zwaan}(2001)}]{martens01}
	{Martens}, P.~C., \& {Zwaan}, C. 2001, \apj, 558, 872
	
	\bibitem[{{Martin}(1973)}]{mar73}
	{Martin}, S.~F. 1973, \solphys, 31, 3
	
	\bibitem[{{Martin}(1998)}]{mar98}
	---. 1998, \solphys, 182, 107
	
	\bibitem[{{McGlasson} {et~al.}(2019){McGlasson}, {Panesar}, {Sterling}, \&
		{Moore}}]{mcglasson19}
	{McGlasson}, R.~A., {Panesar}, N.~K., {Sterling}, A.~C., \& {Moore}, R.~L.
	2019, \apj, 882, 16
	
	\bibitem[{{Moore} {et~al.}(2015){Moore}, {Sterling}, \& {Falconer}}]{moore15}
	{Moore}, R.~L., {Sterling}, A.~C., \& {Falconer}, D.~A. 2015, \apj, 806, 11
	
	\bibitem[{{Panasenco} \& {Martin}(2008)}]{panasenco08}
	{Panasenco}, O., \& {Martin}, S.~F. 2008, Astronomical Society of the Pacific
	Conference Series, Vol. 383, {Topological Analyses of Symmetric Eruptive
		Prominences}, ed. R.~{Howe}, R.~W. {Komm}, K.~S. {Balasubramaniam}, \&
	G.~J.~D. {Petrie}, 243
	
	\bibitem[{{Panesar} {et~al.}(2014){Panesar}, {Innes}, {Schmit}, \&
		{Tiwari}}]{panesar14c}
	{Panesar}, N.~K., {Innes}, D.~E., {Schmit}, D.~J., \& {Tiwari}, S.~K. 2014,
	\solphys, 289, 2971
	
	\bibitem[{{Panesar} {et~al.}(2020){Panesar}, {Moore}, \&
		{Sterling}}]{panesar20}
	{Panesar}, N.~K., {Moore}, R.~L., \& {Sterling}, A.~C. 2020, \apj, 894, 104
	
	\bibitem[{{Panesar} {et~al.}(2017){Panesar}, {Sterling}, \&
		{Moore}}]{panesar17}
	{Panesar}, N.~K., {Sterling}, A.~C., \& {Moore}, R.~L. 2017, \apj, 844, 131
	
	\bibitem[{{Panesar} {et~al.}(2018{\natexlab{a}}){Panesar}, {Sterling}, \&
		{Moore}}]{panesar18a}
	---. 2018{\natexlab{a}}, \apj, 853, 189
	
	\bibitem[{{Panesar} {et~al.}(2016){Panesar}, {Sterling}, {Moore}, \&
		{Chakrapani}}]{panesar16b}
	{Panesar}, N.~K., {Sterling}, A.~C., {Moore}, R.~L., \& {Chakrapani}, P. 2016,
	\apjl, 832, L7
	
	\bibitem[{{Panesar} {et~al.}(2018{\natexlab{b}}){Panesar}, {Sterling}, {Moore},
		{Tiwari}, {De Pontieu}, \& {Norton}}]{panesar18b}
	{Panesar}, N.~K., {Sterling}, A.~C., {Moore}, R.~L., {et~al.}
	2018{\natexlab{b}}, \apjl, 868, L27
	
	\bibitem[{{Panesar} {et~al.}(2019){Panesar}, {Sterling}, {Moore}, {Winebarger},
		{Tiwari}, {Savage}, {Golub}, {Rachmeler}, {Kobayashi}, {Brooks}, {Cirtain},
		{De Pontieu}, {McKenzie}, {Morton}, {Peter}, {Testa}, {Walsh}, \&
		{Warren}}]{panesar19}
	---. 2019, \apjl, 887, L8
	
	\bibitem[{{Parenti}(2014)}]{parenti14}
	{Parenti}, S. 2014, Living Reviews in Solar Physics, 11, 1
	
	\bibitem[{{Pesnell} {et~al.}(2012){Pesnell}, {Thompson}, \&
		{Chamberlin}}]{pensell12}
	{Pesnell}, W.~D., {Thompson}, B.~J., \& {Chamberlin}, P.~C. 2012, \solphys,
	275, 3
	
	\bibitem[{{Raouafi} \& {Stenborg}(2014)}]{raouafi14}
	{Raouafi}, N.-E., \& {Stenborg}, G. 2014, \apj, 787, 118
	
	\bibitem[{{Scherrer} {et~al.}(2012){Scherrer}, {Schou}, {Bush}, {Kosovichev},
		{Bogart}, {Hoeksema}, {Liu}, {Duvall}, {Zhao}, {Title}, {Schrijver},
		{Tarbell}, \& {Tomczyk}}]{scherrer12}
	{Scherrer}, P.~H., {Schou}, J., {Bush}, R.~I., {et~al.} 2012, \solphys, 275,
	207
	
	\bibitem[{{Schmieder} {et~al.}(2008){Schmieder}, {Bommier}, {Kitai},
		{Matsumoto}, {Ishii}, {Hagino}, {Li}, \& {Golub}}]{schmieder08}
	{Schmieder}, B., {Bommier}, V., {Kitai}, R., {et~al.} 2008, \solphys, 247, 321
	
	\bibitem[{{Schou} {et~al.}(2012){Schou}, {Scherrer}, {Bush}, {Wachter},
		{Couvidat}, {Rabello-Soares}, {Bogart}, {Hoeksema}, {Liu}, {Duvall}, {Akin},
		{Allard}, {Miles}, {Rairden}, {Shine}, {Tarbell}, {Title}, {Wolfson},
		{Elmore}, {Norton}, \& {Tomczyk}}]{schou12}
	{Schou}, J., {Scherrer}, P.~H., {Bush}, R.~I., {et~al.} 2012, \solphys, 275,
	229
	
	\bibitem[{{Shen} {et~al.}(2015){Shen}, {Liu}, {Liu}, {Chen}, {Su}, {Xu}, \&
		{Liu}}]{shen15}
	{Shen}, Y., {Liu}, Y., {Liu}, Y.~D., {et~al.} 2015, \apjl, 814, L17
	
	\bibitem[{{Srivastava} {et~al.}(2020){Srivastava}, {Rao}, {Konkol}, {Murawski},
		{Mathioudakis}, {Tiwari}, {Scullion}, {Doyle}, \& {Dwivedi}}]{srivastava2020}
	{Srivastava}, A.~K., {Rao}, Y.~K., {Konkol}, P., {et~al.} 2020, \apj, 894, 155
	
	\bibitem[{{Sterling} {et~al.}(2015){Sterling}, {Moore}, {Falconer}, \&
		{Adams}}]{sterling15}
	{Sterling}, A.~C., {Moore}, R.~L., {Falconer}, D.~A., \& {Adams}, M. 2015,
	\nat, 523, 437
	
	\bibitem[{{Sterling} {et~al.}(2017){Sterling}, {Moore}, {Falconer}, {Panesar},
		\& {Martinez}}]{sterling17}
	{Sterling}, A.~C., {Moore}, R.~L., {Falconer}, D.~A., {Panesar}, N.~K., \&
	{Martinez}, F. 2017, \apj, 844, 28
	
	\bibitem[{{Tandberg-Hanssen}(1995)}]{tand95}
	{Tandberg-Hanssen}, E., ed. 1995, Astrophysics and Space Science Library, Vol.
	199, {The nature of solar prominences}
	
	\bibitem[{{Tian} {et~al.}(2014){Tian}, {DeLuca}, {Cranmer}, {De Pontieu},
		{Peter}, {Mart{\'{\i}}nez-Sykora}, {Golub}, {McKillop}, {Reeves}, {Miralles},
		{McCauley}, {Saar}, {Testa}, {Weber}, {Murphy}, {Lemen}, {Title}, {Boerner},
		{Hurlburt}, {Tarbell}, {Wuelser}, {Kleint}, {Kankelborg}, {Jaeggli},
		{Carlsson}, {Hansteen}, \& {McIntosh}}]{tian14}
	{Tian}, H., {DeLuca}, E.~E., {Cranmer}, S.~R., {et~al.} 2014, Science, 346,
	1255711
	
	\bibitem[{{Tiwari} {et~al.}(2019){Tiwari}, {Panesar}, {Moore}, {De Pontieu},
		{Winebarger}, {Golub}, {Savage}, {Rachmeler}, {Kobayashi}, {Testa}, {Warren},
		{Brooks}, {Cirtain}, {McKenzie}, {Morton}, {Peter}, \& {Walsh}}]{tiwari19}
	{Tiwari}, S.~K., {Panesar}, N.~K., {Moore}, R.~L., {et~al.} 2019, \apj, 887, 56
	
	\bibitem[{{van Ballegooijen} \& {Martens}(1990)}]{balle90}
	{van Ballegooijen}, A.~A., \& {Martens}, P.~C.~H. 1990, \apj, 361, 283
	
	\bibitem[{{Wang} {et~al.}(2018){Wang}, {Liu}, {Li}, {Liu}, {Deng}, {Xu},
		{Jing}, {Wang}, \& {Cao}}]{wang18}
	{Wang}, H., {Liu}, R., {Li}, Q., {et~al.} 2018, \apjl, 852, L18
	
	\bibitem[{{Wyper} {et~al.}(2018){Wyper}, {DeVore}, \& {Antiochos}}]{wyper18}
	{Wyper}, P.~F., {DeVore}, C.~R., \& {Antiochos}, S.~K. 2018, \apj, 852, 98
	
	\bibitem[{{Zirker} \& {Cleveland}(1994)}]{zirker94}
	{Zirker}, J.~B., \& {Cleveland}, F.~M. 1994, \solphys, 153, 245
	
	\bibitem[{{Zirker} {et~al.}(1998){Zirker}, {Engvold}, \& {Martin}}]{zir98}
	{Zirker}, J.~B., {Engvold}, O., \& {Martin}, S.~F. 1998, \nat, 396, 440
	
\end{thebibliography}

\end{document}